\documentclass[twocolumn,showpacs,prl]{revtex4}% Physical Review B
\usepackage{graphicx}

\begin{document}

\title{Performance engineering of semiconductor spin qubit systems}

\author{Ramin M. Abolfath$^{1,2}$, Thomas Brabec$^{1}$}

\affiliation{
$^{1}$University of Ottawa, Physics Department 150 Louis Pasteur, Ottawa, ON, K1N 6N5, Canada\\
$^{2}$School of Natural Sciences and Mathematics, University of Texas at Dallas, Richardson, TX 75080}
\date{\today}

\begin{abstract}
The performance of a quantum computation system is investigated, with qubits represented by magnetic impurities in coupled quantum dots filled with two electrons. Magnetic impurities are electrically manipulated by electrons. The dominant noise source is the electron mediated indirect coupling between magnetic impurities and host spin bath. As a result of the electron mediated coupling, both noise properties and the time needed for elementary gate operations, depend on controllable system parameters, such as size and geometry of the quantum dot, and external electric and magnetic fields. We find that the maximum number of quantum operations per coherence time for magnetic impurities increases as electron spin singlet triplet energy gap decreases. The advantage of magnetic impurities over electrons for weak coupling and large magnetic fields will be illustrated.
\end{abstract}

\maketitle
Semiconductor nanostructures~\cite{Jacak:book} are a promising host material for the realization of quantum
computers~\cite{Galindo2002:RMP} because of the well developed production technologies and because of the
potential for scaling to multi qubit systems. The greatest bottleneck for realizing semiconductor quantum devices
is the high degree of noise present in the host material.

Recently substantial progress has been made in the experimental demonstration of electron spin coded qubits in lateral double QDs~\cite{QDMexperiment1}, first theoretically proposed by Loss and DiVincenzo~\cite{Loss1998:PRB}. This system exhibits promising properties for quantum information processing, as it can be controlled electrically like charge qubits, however offers the longer decoherence times of spin qubits. Nevertheless, dephasing in spin coded qubits due to nuclear hyperfine interaction still presents the major hurdle for realizing a scalable quantum computer \cite{Cywiski2009:PRL,coish2009:PSSB}.

Another seminal idea of realizing quantum computation in semiconductors is by the use of a hybrid qubit, first proposed by Kane in a $^{31}$P doped silicon host \cite{Cory1998:Nature,Kane:1998:Nature,Galindo2002:RMP}. In silicon, $^{31}$P is a positive dopant that is transformed into a positively charged nucleus and a loosely bound valence electron. Whereas, the qubit is represented by the nuclear spin of $^{31}$P, electric gate operations on the electron are used for manipulation and diagnostics of the qubit. The technological implementation of this system is more challenging, mostly as a result of the significantly smaller dimensions of the qubits.

\begin{figure}
\begin{center}
\vspace{1cm}
\includegraphics[width=0.98\linewidth]{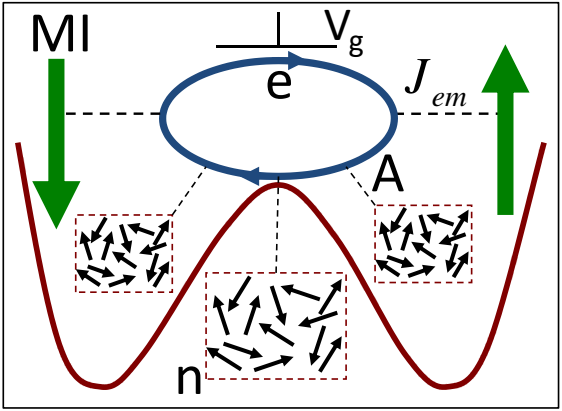}
% Path:
\caption{Schematics of two spin qubit unit in a DQD considered here. Each spin qubit is represented by a MI.
Interaction between qubits is mediated by the two electrons; the electron MI coupling coefficient is denoted by $J_{\rm em}$. As interaction of MIs with nuclear spin bath is negligible, the dominant noise source is electron mediated coupling between MIs and nuclear spin bath; the electron nuclear spin bath coupling coefficient is given by $A$. The interaction strength between the MIs and between MI and nuclear spin bath depends on the singlet-triplet electron energy gap $\Delta_e$, which can be controlled by external electric gate voltage ($V_g$), magnetic field ($B$), and shape/size of the DQD. This allows active noise engineering and optimization of the two qubit system.}
\label{fig1}
\end{center}
\end{figure}

We suggest and analyze here a qubit that is a mixture of the spin coded \cite{Loss1998:PRB} and of the hybrid
qubits \cite{Cory1998:Nature,Kane:1998:Nature} discussed above. The proposed device consists of a QD containing electrons and a neutral dopant acting as a magnetic impurity (MI). Similar to the original hybrid qubit, the MI represents the qubit, and the electrons are used for manipulating the qubit. The use of MIs alone for spin-based quantum devices and its noise spectroscopy has been investigated recently~\cite{QDMexperiment2,Fernandez2004:PRL,Nussinov2003:PRB}. Here, we confine our analysis to MIs with zero nuclear spin, [e.g., $^{56}$Fe/$^{160}$Gd doped in II-VI/IV-VI materials], to exclude coupling between spins of the electron and nucleus of the MI. As shown in Fig.\ref{fig1}, the MI is localized in space and its direct interaction with the host spin bath is negligible. Therefore, the decoherence of the qubit is determined by the electron mediated coupling of the MI to the spin bath.

Our configuration has two advantages over the original qubit systems. First, it allows greater design flexibility, in particular with regard to the size of the confining potential. This will facilitate the actual technological realization of a hybrid qubit. Second, MI-electron coupling and electron mediated MI-spin bath coupling show particular dependence on the electron spin singlet triplet energy gap $\Delta_{\rm e}$, that allows forming stable qubit over a range of QD confining potentials, and the external electric and magnetic fields. This system corresponds to the electron spin coded qubit system investigated in Ref. \cite{Loss1998:PRB} and makes a comparison possible. The performance of our quantum computing setup is measured by the maximum number of operations $N_i = \tau_i / T_i$ with $i = e,m$ for electrons and MI qubits, respectively. Here, $\tau$ is the coherence time and $T$ is the time required for an elementary gate operation. In this work we illustrate that the quantum performance of the spin coded qubit based on MIs increases by the external electric and magnetic fields. This is in contrast to the electron spin coded qubit that its performance is suppressed rapidly by the inter-dot coupling.
% ================================ intro end ==============================================

We represent the QD system by the Hamiltonian $H=H_{\rm e} + H_{\rm m} + H_{\rm n} + H_{\rm in}$, which contains
electrons ($H_{\rm e}$), MIs ($H_{\rm m}$), host semiconductor nuclei ($H_{\rm n}$), and their interactions
($H_{\rm in}$). In our model we consider e-MI and e-n interaction, i.e. $H_{\rm in}=H_{\rm em}+H_{\rm en}$.
The direct interaction between MI electrons and semiconductor host nuclei are neglected, as d- and f-electrons
are highly localized in space. Further, our analysis is confined to MIs with zero nuclear magnetic moment, which
do not interact with the MI electrons. As a result, dipole-dipole interaction between the nuclear spins of MI
and host nuclei has no influence on our system.

Electrons confined in quasi-two-dimensional quantum dots in a uniform perpendicular magnetic field can be described
by the effective mass Hamiltonian
\begin{equation}
H_{\rm e} = \sum_{i=1}^N  \left( T_i + Z_{i} \right) + \frac{e^2}{2\varepsilon}\sum_{i \neq j}\frac{1}{|\vec{r}_i
- \vec{r}_j|},
\label{hame}
\end{equation}
where $T = 1/(2m^*) \left( (\hbar/i) \vec{\nabla} + (e/c) A(\vec{r})\right)^2 + V(\vec{r})$ is the single electron
Hamiltonian in an external magnetic field $\vec{B} = B \hat{z}$, perpendicular to the plane of the confining potential. Here $(\vec{r})=(x,y,z)$ describes the electron position, $V(\vec{r})$ denotes the quantum dots confining potential, and $A(\vec{r})=(1/2) \vec{B}\times\vec{r}$ is the vector potential. Further, $m^*$ is the conduction-electron effective mass, $-e$ is the electron charge, and $\varepsilon$ is the host semiconductor dielectric constant. Finally, $Z_{i}= (1/2) g_{\rm e} \mu_{\rm b} S_{zi} B$ determines the Zeeman spin splitting, $g_{\rm e}$ is the electron g-factor in host semiconductor, $\mu_{\rm b}$ refers to the Bohr magneton, and $S_{zi}$ represents the $z$-Pauli matrix of electron $i$.

The single particle eigenvalues ($\epsilon_{\alpha\sigma}$) and eigenvectors ($\varphi_{\alpha\sigma}$) are 
calculated by 
discretizing $T+Z$ in real space, and diagonalizing the resulting matrix. By using the creation (annihilation) 
operators $c^\dagger_{\alpha\sigma}~(c_{\alpha\sigma})$ for an electron in a non-interacting single-particle (SP) state 
$|\alpha,\sigma\rangle$, the Hamiltonian of an interacting system in second quantization can be written as
\begin{eqnarray}
H_e&=&\sum_{\alpha}\sum_{\sigma} \epsilon_{\alpha\sigma} c^\dagger_{\alpha\sigma} c_{\alpha\sigma} \nonumber \\ &&
+ \frac{1}{2}\sum_{\alpha\beta\mu\nu}\sum_{\sigma\sigma'} V_{\alpha\sigma,\beta\sigma',\mu\sigma',\nu\sigma} 
c^\dagger_{\alpha\sigma} 
c^\dagger_{\beta\sigma'}
c_{\mu\sigma'} c_{\nu\sigma},
\label{multiparticle}
\end{eqnarray}
where the first term represents the single particle Hamiltonian and
${V}_{\alpha\sigma,\beta\sigma',\mu\sigma',\nu\sigma} =
\int d\vec{r} \int d\vec{r'}
{\varphi}^*_{\alpha\sigma}(\vec{r})
{\varphi}^*_{\beta\sigma'}(\vec{r'})
\frac{e^2}{\varepsilon|\vec{r}-\vec{r'}|}
{\varphi}_{\mu\sigma'}(\vec{r'})
{\varphi}_{\nu\sigma}(\vec{r})$,
is the two-body Coulomb matrix element.

The Hamiltonian for the MIs accounts for MI-MI direct exchange interaction and MI-Zeeman coupling,
\begin{equation}
H_{\rm m} = \sum_{j,j'=1}^M J_{jj'} \vec{M}_j \cdot \vec{M}_{j'} + \sum_{j} g_{\rm m} \mu_{\rm b} M_{zj} B,
\label{hamm}
\end{equation}
where $J_{jj'}$ is the direct MI-MI antiferromagnetic (AFM) coupling, $g_{\rm m}$ is the MI g-factor, and $M_{zj}$ is the z-component of the MI spin operator.

The nuclear-nuclear direct dipole interaction in the host semiconductor is neglected. The nuclear Hamiltonian is given by the Zeeman coupling term,
\begin{equation}
H_{\rm n} = \sum_{l=1}^L g_{\rm n} \mu_{\rm b} I_{zl} B,
\label{hamn}
\end{equation}
where $g_{\rm n}$ is the nuclear g-factor and $I_{zl}$ is the z-component of nuclear spin operator.

The e-MI exchange interaction is modeled by
\begin{equation}
H_{\rm em}=- J_{\rm em} \sum_{i,j}\vec{S}_i \cdot\vec{M}_j \delta({\bf r}_i - {\bf R}_j),
\label{hamem}
\end{equation}
with $J_{\rm em}$ the exchange coupling between electron spin $\vec{S}_i$ at ${\bf r}_i$ and impurity spin $\vec{M}_j$ located at the position ${\bf R}_j$~\cite{Fernandez2004:PRL}. In second quantization it can be written as
\begin{eqnarray}
H_{em} &=& - \sum_{\alpha\beta}\sum_I \frac{J_{\alpha\beta}({\bf R}_j)}{2}
[M_{zj}
(c^\dagger_{\alpha\uparrow}c_{\beta\uparrow}-
c^\dagger_{\alpha\downarrow}c_{\beta\downarrow}) \nonumber \\ &&
+ M_j^+ c^\dagger_{\alpha\downarrow}c_{\beta\uparrow}
+ M_j^- c^\dagger_{\alpha\uparrow}c_{\beta\downarrow}],
\label{Hex}
\end{eqnarray}
where $J_{\alpha\beta}({\bf R}_j) = J_{em}\varphi^*_{\alpha}({\bf R}_j) \varphi_{\beta}({\bf R}_j)$. 
Similarly, we describe the electron - nuclear spin bath hyperfine interaction by
\begin{equation}
H_{\rm en}= \sum_{i,l} \tilde{A}_l \vec{S}_i \cdot \vec{I}_l \delta({\bf r}_i - {\bf R}_l) 
\label{hamen}
\end{equation}
with $\tilde{A}_l = (16\pi/3) \mu_b \mu_l/I_l$ the isotropic (Fermi contact) part of the electron-nucleus hyperfine interaction ~\cite{LandauLifshitzQM:book,coish2009:PSSB,Merkulov2002:PRB}. Here $\mu_l$ and ${\bf R}_l$ are magnetic moment, and position of the $l$th nucleus and sum goes over all nucleus in the lattice. In second quantization it can be written as
\begin{eqnarray}
H_{en} &=& - \sum_{\alpha\beta}\sum_l \frac{A_{\alpha\beta}({\bf R}_l)}{2}
[I_{zl}
(c^\dagger_{\alpha\uparrow}c_{\beta\uparrow}-
c^\dagger_{\alpha\downarrow}c_{\beta\downarrow}) \nonumber \\ &&
+ I_l^+ c^\dagger_{\alpha\downarrow}c_{\beta\uparrow}
+ I_l^- c^\dagger_{\alpha\uparrow}c_{\beta\downarrow}],
\label{Hex}
\end{eqnarray}
where $A_{\alpha\beta}({\bf R}_l) = \tilde{A}_l\varphi^*_{\alpha}({\bf R}_l)
\varphi_{\beta}({\bf R}_l)$. 
Finally
\begin{eqnarray}
H_{\rm int}=-\frac{1}{2}\sum_{\alpha\alpha'}\sum_{\sigma\sigma'}
\vec{Q}_{\alpha\alpha'}\cdot\tau_{\sigma\sigma'}
c^\dagger_{\alpha\sigma} c_{\alpha'\sigma'},
\end{eqnarray}
where $\vec{Q}_{\alpha\alpha'}=\sum_I J_{\alpha\alpha'}(\vec{R}_j)\vec{M}_j
-\sum_n A_{\alpha\alpha'}(\vec{R}_l)\vec{I}_l$.

From the total Hamiltonian $H$ an effective Hamiltonian is obtained by tracing over the degrees of freedom of the electron wavefunction and by taking the interaction term $H_{\rm in}$ into account to second order of perturbation theory, which yields
\begin{equation}
H_{\rm eff} = H_{\rm m} + H_{\rm n} + \sum_{\rm x} {|\langle \Psi_{\rm x}| H_{\rm in} |\Psi_{\rm g} \rangle|^2 \over
E_{\rm g} - E_{\rm x} }.
\label{hameff1}
\end{equation}
Here we limit our calculation to a two electron and two MI system in a DQD. The two-electron wavefunction is confined to the Hilbert sub-space constructed from the bonding and anti-bonding (HOMO, LUMO) one-electron orbitals of the DQD, $\varphi_{\pm}$. Below the magnetic field corresponding to spin singlet-triplet transition, this results in six basis functions of two-electron, a spin singlet ($S_0$) ground state $\Psi_{\rm g}$ that can be expressed as superposition 
of $\varphi_+(\vec{r}_1) \varphi_+(\vec{r}_2) |S_0 \rangle$ and $\varphi_-(\vec{r}_1) \varphi_-(\vec{r}_2)
|S_0 \rangle$ with binding energy $E_{\rm g}$, and five excited states $\Psi_{\rm x}$ with energy $E_{\rm x}$,
consisting of three degenerate first excited triplet states, and two higher excited singlet 
states~\cite{Abolfath2009:PRB}.

Calculating the matrix elements in Eq. \ref{hameff1} yields the effective Hamiltonian
\begin{equation}
H_{\rm eff} = H_{\rm m} + H_{\rm n} + H_{\rm mm} + H_{\rm mn},
\label{hameff2}
\end{equation}
where $H_{\rm mm} = \sum_{j,j'} \Delta_{j j'} \vec{M}_j \cdot \vec{M}_{j'}$ is the electron mediated
(RKKY-type \cite{RKKY}) interaction between the MIs, and $H_{mn} = \sum_{j,l} \Delta_{jl} \vec{I}_{l} \cdot \vec{M}_j$ is the electron mediated interaction between MIs and nuclear spin bath; the electron mediated interaction between host nuclear spins $H_{\rm nn}$ is neglected. Here, $\Delta_{j j'} = - \gamma^2 J^2_{em} U(\vec{R}_j,\vec{R}_{j'})/(2 \Delta_e)$ and $\Delta_{j,l} = \gamma^2 \tilde{A} J_{em} U(\vec{R}_j,\vec{R}_{l})/\Delta_e$.
Further, $\gamma = \alpha_+ -\alpha_-$, where $\alpha_+$, and $\alpha_-$ are the coefficients of the 
two-electrons ground state that is expressed as linear combination of bonding-antibonding in two level model
$\Psi_g(\vec{r}_1, \vec{r}_2) =
\left[\alpha_+\varphi_+(\vec{r}_1) \varphi_+(\vec{r}_2)
     + \alpha_-\varphi_-(\vec{r}_1) \varphi_-(\vec{r}_2) \right] |S_0\rangle$,
$\Delta_e$ is the two electron singlet-triplet splitting, $\tilde{A} = 1 / L \sum_{l=1}^L \tilde{A}_l$, and $U(\vec{R}_j,\vec{R}_{\lambda}) = \varphi_{+}(\vec{R}_j) \varphi_{-}(\vec{R}_j) \varphi_{+}(\vec{R}_{\lambda}) \varphi_{-}(\vec{R}_{\lambda})$ with $\lambda=j',l$.
% The tunable parameter $\Delta_{jl}$ represents the strength of the effective interaction between MI and host
% nuclear spins.

\begin{figure}
\begin{center}
\vspace{1cm}
\includegraphics[width=0.98\linewidth]{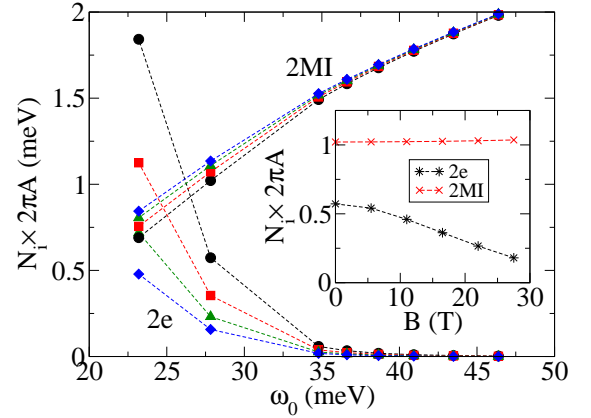}
%\includegraphics[width=0.98\linewidth]{Fig2b.pdf}
% Path:
\caption{
$N_{\rm i}$, the maximum number of elementary gate operations within the coherence-time normalized by factor $2\pi A$ 
($A\equiv A_e\approx A_m$) in meV in a Fe:ZnSe DQD (2MI), compared with ZnSe DQD filled with two-electrons (2e) versus parabolic strength of confining potential of each dot, $\omega_0$, for various gate voltages, $V_g=110$ (circles), $V_g=154$ (squares),
$V_g=198$ (triangles), and $V_g=242$ (diamonds) in meV. $N_{\rm i}$ as a function of external magnetic field for $\omega_0=27.5$ meV and $V_g=110$ meV is shown in the inset.
}
\label{fig2}
\end{center}
\end{figure}

In the following we use the effective Hamiltonian \ref{hameff2} to calculate the decoherence time of the DQD with two MIs and two electrons, where each MI represents a spin qubit. The result is compared to a two qubit system realized by two electrons in a DQD. The calculation is performed by using the quasi-static bath
approximation~\cite{Merkulov2002:PRB,Dobrovitski2002:PRE,Melikidze2004:PRB}, where the host nuclear spins are
approximated by a random magnetic field $\vec{B}_{\rm n}$ with a Gaussian distribution.
%\begin{equation}
%w(B_{\rm n}) = {1\over \pi^{3/2} \Delta_{\rm B}^3} \exp \left( -{B_n^2 \over \Delta_B^2} \right).
%\label{randomb}
%\end{equation}
In this limit the two-electron and two-MI nuclear bath Hamiltonian are given by
\begin{equation}
H_{\rm kn} = \sum_{i=1}^2 g_{\rm k} \mu_{\rm b} \vec{B}_{\rm n} \cdot \vec{K}_i,
\label{hen}
\end{equation}
where $\rm k = e,m$ and $\vec{K} = \vec{S}, \vec{M}$ for electrons and MI, respectively. The coherence time is obtained by solving the equation of motion for $\vec{K}_1$ and $\vec{K}_2$ with initial state $\mid \uparrow \downarrow \rangle$ and by averaging over the Gaussian magnetic field distribution. From that we obtain 
$\langle B_{\rm e} \rangle=\langle B_{\rm m} \rangle=0$, 
$\langle B^2_{\rm e} \rangle = 1/(g_{\rm e} \mu_{\rm b})^2
\sum_l I_l (I_l + 1) \tilde{A}^2_l |\varphi_{+}({\bf R}_l)|^4$, and
$\langle B^2_{\rm m} \rangle = 1/(g_{\rm m} \mu_{\rm b})^2
\sum_l I_l (I_l + 1) \tilde{A}^2_l |\varphi_{+}({\bf R}_l)\varphi_{-}({\bf R}_l)|^2$.
From there an effective Zeeman splitting $\tilde{\Delta}_{\rm k}=(2\langle B^2_{\rm k} \rangle/3)^{1/2}$ 
is calculated, hence $\tau_{\rm k}=\hbar/(g_{\rm k} \mu_{\rm b} \tilde{\Delta}_{\rm k})$. 
Assuming $I_l=1/2$ we find the spin relaxation time 
$\tau_{\rm e}=\hbar/(2A_e)$ and $\tau_{\rm m}=C \hbar/(2A_m)$. Here  
$A_e =( \sum_l \tilde{A}^2_l |\varphi_{+}({\bf R}_l)|^4 )^{1/2}$, 
$A_m =( \sum_l \tilde{A}^2_l |\varphi_{+}({\bf R}_l)\varphi_{-}({\bf R}_l)|^2 )^{1/2}$, and
$C=\Delta_{\rm e}/[\gamma^2 J_{\rm em}\Lambda(\vec{R}_1, \vec{R}_2)]$ is the RKKY correction to the MI coherence time, stems from the MI-nuclear-spin interaction mediated by electrons. Here $\Lambda(\vec{R}_1, \vec{R}_2) = \sum_{j=1}^2 |\varphi_+(\vec{R}_j)\varphi_-(\vec{R}_j)|$ describes the spatial dependence of the e-MI exchange interaction, a parameter that depends on the electron envelop wavefunction at the MI positions $\vec{R}_j$. Note that in the limit 
of zero inter-dot tunneling, $A_e=A_m$. The ratio of MI and electron coherence times can be calculated as
\begin{equation}
{\tau_{\rm m} \over \tau_{\rm e}} = \frac{\Delta_{\rm e}}{\gamma^2 J_{\rm em}\Lambda(\vec{R}_1, \vec{R}_2)}
\frac{A_e}{A_m}.
\label{timer}
\end{equation}

The performance of a quantum computing setup is given by the maximum number of operations $N_i = \tau_i / T_i$ with $i = e,m$ for electrons and MI qubits, respectively. Here, $\tau$ is the coherence time and $T$ is the time required for the elementary gate operations. Our system is compared to the original proposal in Ref. \cite{Loss1998:PRB}, where an XOR gate control in a two-electron DQD is analyzed. The elementary gate operations needed for the XOR gate are: (i) a correlated spin swap from $\mid \uparrow \downarrow \rangle \rightarrow \mid \downarrow \uparrow \rangle$, where the first and second position refers to the left and right dot, respectively, and (ii) single qubit operations with an external pulsed magnetic field. The second operation has to be done within the time of one spin swap. Therefore, the time for one XOR gate is determined by the correlated spin swap time $T$.

The MI and electron spin swap times are given by $\Delta^{-1}_{\rm m}$, and $\Delta^{-1}_{\rm e}$ modulus $\pi\hbar$, assuming that the gate voltage and therewith exchange coupling is controlled by a square pulse~\cite{Loss1998:PRB}. Here, $\Delta_{\rm m} \equiv \Delta_{jj'}$ is the coupling coefficient between the two MIs. As a result, $N_{\rm e}=\Delta_{\rm e}/(2\pi A_e)$, $N_{\rm m}= [J_{\rm em} U/ (2\Lambda)]/(2\pi A_m)$ where $U= \varphi_{+}(\vec{R}_1) \varphi_{-}(\vec{R}_1) \varphi_{+}(\vec{R}_2) \varphi_{-}(\vec{R}_2)$ and $\vec{R}_1$, $\vec{R}_2$ are MI coordinates. The ratio of the maximum number of elementary operations (per coherence time) is given by $N_{\rm m} / N_{\rm e} = (\tau_{\rm m} / \tau_{\rm e}) (\Delta_{\rm m}/\Delta_{\rm e})$, which finally gives
\begin{equation}
{N_{\rm m} \over N_{\rm e}} = {J_{\rm em} \over \Delta_{\rm e}} \frac{U}{2\Lambda} \frac{A_e}{A_m}.
\label{final}
\end{equation}
In Eq.~(\ref{final}), $N_{\rm m} / N_{\rm e} \propto 1/\Delta_{\rm e}$. Unlike the other parameters in Eq.~(\ref{final}), $\Delta_{\rm e}$ decays to zero very rapidly by increasing the external magnetic field and inter-dot energy barrier that lowers the inter-dot coupling. Therefore one expects to observe decay in performance of the electron spin coded qubit due to variations in $B$ and $V_g$. Unlike the electrons, MIs show a robust increase in their quantum operation performance. To gain the optimum performance of MIs over electrons we employ a numerical calculation based on exact diagonalization of Eq.(\ref{hame}) from which the input parameters for Eq.(\ref{final}) are obtained~\cite{Abolfath2009:PRB}. The DQD is chosen to be double Gaussian along the axis and parabolic in the perpendicular direction and each MI is centered at one of the QDs. We perform our calculation for three different materials, Gd:PbTe, Fe:CdSe, and Fe:ZnSe where they show $\tau_{\rm m}/\tau_{\rm e} \approx 300, 20, 2$ respectively at $V_g=154$ meV and $B=0$. The values for $J_{\rm em}$ are adopted from Ref.~\cite{Dietl1994:PRB}. In this range of parameters the coupling between MIs and nuclear spins ($\Delta_{j,l}$) is optimized to be weak to maximize $\tau_{\rm m}/\tau_{\rm e}$. However with decreasing $\Delta_{j,l}$, the coupling between two MIs ($\Delta_{\rm m}$) lowers and as a result the time required for fundamental gate operations becomes longer. Thus, the gain in coherence time ($\tau_{\rm m}/\tau_{\rm e}$) is offset by a loss in gate operation times ($\Delta_{\rm e}/\Delta_{\rm m}$). To maximize $N_{\rm m}/N_{\rm e}$ we search for a range of parameters that allows simultaneous maximization of $\tau_{\rm m}/\tau_{\rm e}$ and $\Delta_{\rm m}/\Delta_{\rm e}$.

In Fig. \ref{fig2} we show the results for Fe:ZnSe. $N_{\rm m}$ and $N_{\rm e}$ normalized to $2\pi A_e$ are plotted 
as a function of  parabolic confining strength $\omega_0$ for various values of the gate voltage $V_g$, and the 
external magnetic field $B$ (inset). Within numerical parameters considered in this calculation we found 
$A_e\approx A_m$. We observe that $N_{\rm m}/N_{\rm e}$ increases with increasing $B$, $V_g$ (inter-dot energy barrier) and $\omega_0$ (tighter confinement). Within the parameter range considered here, a maximum performance increase of about three orders of magnitude can be achieved over the electron spin coded qubit.
It is important to mention that even higher increases of $N_{\rm m}/N_{\rm e}$ might be achievable. The maximum $B$ and $V_g$ values used for our optimization had to be limited to the range of validity of our two level model. For increasing values of $B$ and $V_g$ the system approaches the singlet triplet transition point at which $\Delta_{\rm e} \rightarrow 0$ and  $N_{\rm m}/N_{\rm e} \propto 1/\Delta_{\rm e} \rightarrow \infty$. In this limit our approximation based on the two level model fails, and a more exact analysis becomes necessary. This will be studied in more detail in a follow-up work.
Further, the smallest realizable quantum dot size is around 5 nm corresponding to the maximum $\omega_0$ in Fig. \ref{fig2}. However, an extrapolation of the numerical results to the atomic scale indicates performances gains $N_{\rm m}/N_{\rm e}$ of more than five orders of magnitude. In this limit, our system becomes comparable to Kane's proposal~\cite{Kane:1998:Nature}, which demonstrates its favorable performance properties. The advantage of our system is that a compromise can be found between optimizing performance and accommodating technological limitations.

In conclusion the qubit system investigated here opens the possibility for noise and performance optimization. We have found that a combination of active and passive optimization is necessary to obtain appreciable
improvements; our analysis of the hybrid MI/electron qubit predicts a performance gain of at least three orders of magnitude over electron spin coded qubits in the limit of small inter-dot coupling.

%%%%%%%%%%%%%%%%%%%%%%%%%%%%%%%%%%%%%%%%%%%%%%%%%%%%%%%%%%%%%%%%

%%%%%%%%%%%%%%%%%%%%%%%%%%%%%%%%%%%%%%%%%%%%%%%%%%%%%%%%%%%%%%%%%%%%%%%%%%%%%%%

\end{document}